\documentclass[11pt]{article}

\usepackage{graphicx}
\usepackage{indentfirst}
\usepackage{booktabs}
\usepackage{fullpage}
\usepackage{array}
\usepackage{hyphenat}
\usepackage{setspace}
\usepackage{color}
\usepackage{soul}
\usepackage{epsfig}
\usepackage{subfigure}
\usepackage{url}
\usepackage{algorithm}
\usepackage{algpseudocode}

\begin{document}

\title{Hierarchical Range Sectoring and Bidirectional Link Quality Estimation for On-demand Collections in WSNs}

\author{V\'\i{}ctor Valls, Jos\'e  Luis S\'anchez, Cristina Cano, Boris Bellalta, Miquel Oliver\\
Department of Information Technologies and Communications\\
Universitat Pompeu Fabra\\
Corresponding author: \texttt{victor.valls@upf.edu}}
\date{}
\maketitle

\begin{abstract}
The paper presents two mechanisms for designing an on-demand, reliable and efficient collection protocol for Wireless Sensor Networks. The former is the \textit{Bidirectional Link Quality Estimation}, which allows nodes to easily and quickly compute the quality of a link between a pair of nodes. The latter, \textit{Hierarchical Range Sectoring}, organizes sensors in different sectors based on their location within the network. Based on this organization, nodes from each sector are coordinated to transmit in specific periods of time to reduce the hidden terminal problem. To evaluate these two mechanisms, a protocol called HBCP (Hierarchical-Based Collection Protocol), that implements both mechanisms, has been implemented in TinyOS 2.1, and evaluated in a testbed using TelosB motes. The results show that the HBCP protocol is able to achieve a very high reliability, especially in large networks and in scenarios with bottlenecks.
\end{abstract}


\section{Introduction}

Wireless Sensor Networks (WSNs) are formed by small devices that can be located in a wide range of scenarios, and usually in non-easily accessible places: underground for pressure measurements, throughout the forest to detect fires or inside buildings to take environmental measures, among many others. Today's trend to monitor everything, everywhere and at any time leads to a wide range of WSNs applications, that have multiple and diverse requirements in terms of reliability, scalability and network life-time \cite{akyildiz2004wireless}. 

Sensors are basically composed by a micro-controller (MCU), a transceiver to communicate with other devices, a sensorboard to sense environmental data and a battery to power the whole device. There are many different types of sensor platforms, composed by different hardware components, but all of them have in common the fact that they need to be powered by a battery of limited capacity. Due to this factor, and the difficulty to access the sensors location to replace the battery, it is essential to find mechanisms that maximize its life-time while still achieving the required network performance. 

There is an extensive literature regarding the maximization of the network life-time, from MAC protocols like the B-MAC \cite{polastre2004versatile} or S-MAC \cite{ye2004medium} that try to reduce the energy consumption by allowing sensor nodes to go to sleep and wake-up periodically, to network protocols as LEACH \cite{heinzelman2000energy}, which has the aim to balance the energy consumption during data collections to maximize the life-time of the whole network. However, the performance of a given protocol varies considerably depending on the type of topology and the application the network has been deployed for. In general, there is no protocol stack that perfectly fits all the scenarios and applications, and the election directly depends on the network purpose and the environment characteristics. 

Regarding data collection protocols for WSNs, they can be classified into two groups based on the trigger point of view \cite{akkaya2005survey}. On the one hand, the ones where nodes individually decide whether to send data based on an environmental lecture, and on the other hand, on-demand protocols, like the Direct Diffusion \cite{intanagonwiwat2003directed}, where the sink triggers a query to collect data. The first set of protocols are suitable for monitoring applications, where the others are better, among others, for metering-like applications that have low collection periodicity. In the latter case, the nodes only need to be active during a short period of time, few minutes every month, and then they can remain in a low-consumption state the time they do not need to be active. In this kind of WSNs, it is possible to have large network life-times in the order of several years with standard batteries. 

This paper focuses on a scenario where the sink triggers a query to collect one reading from each sensor in a network, i.e. on-demand collections. We present two new mechanisms called \textit{Hierarchical Range Sectoring} (HRS) and  \textit{Bidirectional Link Quality Estimation} (BLQE). In addition, a new protocol called Hierarchical-Based Collection Protocol (HBCP) is defined, which apart from these two new mechanisms, also includes all the required functionalities to perform efficient on-demand data collections in WSNs. The HBCP design requirements are: $1$) collections must be carried out as fast as possible to minimize the time that the network is awake and thus, save energy, $2$) it has to be able to operate satisfactorily in different topologies and with different workloads, independently of the network or application requirements and $3$) a high next hop and end-to-end delivery rate is required (i.e., close to $99$\%). To assess the performance of the HBCP protocol, and of the HRS and BLQE mechanisms 
in particular, the HBCP has been implemented in TinyOS 2.1 and experimentally evaluated in a testbed.

The paper is divided as follows. Firstly, Section \ref{Sec:motivation} describes the main problems that WSNs suffer for the type of network we focus on. Secondly, Section \ref{mechanisms} presents a general description of the HRS and BLQE mechanisms, and afterward, in Section \ref{description}, it is described in detail how the protocol works throughout all the collection process. Section \ref{software} presents how the HBCP has been implemented in TinyOS 2.1, and in Section \ref{evaluation} the testbed evaluation and the main results are described. Finally, Section \ref{Sec:RelWork} compares the HBCP with other existing protocols, and Section \ref{conclusions} concludes the article.

\section{Open Challenges for an On-demand Data Collection Protocol for WSNs} 
\label{Sec:motivation}

In this section the main challenges that an on-demand data collection protocol for WSNs has to successfully handle are described. These are basically the energy consumption and the scalability. In addition, for on-demand multihop data collections that aim to perform the collection as fast as possible and with minimum energy expenditure, the hidden terminal problem needs to be taken into account as it may significantly harm the network reliability.

\subsection{Energy Consumption} 

The sensor's components that affect the most the battery consumption are the MCU and the transceiver, and depending on the used platform, which defines these components, the overall consumption may vary considerably. These two components are key for the correct operation of the sensor, despite they do not need to be active if they do not have a task to perform. Some platforms have the possibility to turn off the radio and put the MCU into sleep mode in order to save energy, so the nodes only wake up when they have some task to perform. Moreover, the faster they do their task, the sooner they can go back to sleep and save energy. However, determining whether a node has to be active is not trivial, especially if they belong to a network where nodes have to forward data from other nodes. 

Furthermore, regarding the energy consumption of the radio, the behavior of one node directly affects the life-time of the neighboring nodes. In some of the  current transceivers for WSNs, like the CC2420 \cite{instruments2007cc2420}, the energy consumed when receiving a byte is equal or even a slightly higher than transmitting one. Because of this, and the fact that for every transmitted packet the neighboring nodes will overhear it, it is essential to reduce the number of packets sent, especially when the density of the nodes is high. 

\subsection{Scalability} 
Scalability is another problem that multihop WSNs face, and it can appear due to the number of hops to reach the sink or due to the density of nodes. Both of them directly affect the network reliability, as the number of hops grows the probability of an intermediary link break down increases, and as a result the delivery probability drops to 0 ($\lim_{h \to \infty} (1-p)^h = 0 $ where $(1-p)$ is the probability of a successful transmission and \emph{h} the number of hops). Either way with the density of nodes, the larger number of nodes in a certain area, the higher the probability of collisions. To check if the medium is free to transmit is responsibility of the MAC layer, however, to base the decision to transmit on carrier sensing does not necessarily mean that the medium is idle at reception, due to the hidden terminal problem. 

\subsection{Multihop Communications and Hidden Terminals} 
\label{terminal}
The hidden terminal problem can be divided in two cases: 1) when nodes want to send data to the same node and they cannot sense each other, and 2) when the transmission of one node indirectly affects the on-going transmission between two other nodes. With random medium access protocols, both cases have no solution despite being mechanisms like the RTS-CTS \cite{fullmer1997solutions} that try to minimize the impact of collisions. The main difference between the two cases, is that applying a time scheduling to alleviate this problem in the second case is less complex than the one needed for the first case. In the first case each node must be individually scheduled, while in the second case it is possible to create a schedule for groups of nodes. In Section \ref{subsection:RHC} is explained how nodes can be isolated in \emph{Sectors} in order to reduce this kind of collisions.

\section{HBCP Mechanisms} 
\label{mechanisms}

In the following sections, the mechanisms that the HBCP implements are described. The two main mechanisms are the Hierarchical Range Sectoring and the Bidirectional Link Quality Estimator which define the sectors and the datapaths respectively. The other presented mechanisms, Randomized Transmissions and Data Aggregation, are required to implement an efficient on-demand data collection protocol for WSNs.

\subsection{Hierarchical Range Sectoring (HRS)}
\label{subsection:RHC}

HRS consists of grouping nodes in sectors according to their minimum number of hops to reach the sink with good quality links (see Section \ref{subsection:LQE}), with the aim to allow nodes to transmit simultaneously with a low collision probability. 
To better understand the concept of \emph{Sectoring} see the following example: if a node sends a broadcast packet at a fixed transmission power, all the nodes that receive that packet with enough quality are at distance 1 from the initial node. Then, if each of those nodes also send a broadcast packet, the new nodes that receive those packets will be classified as distance 2; and so on for the following sectors. This assures that generally, if the medium conditions do not change, non-continuous sectors would not be able to sense each other. 

\begin{figure}[ht!]
  \centering
    \includegraphics[width=0.35\textwidth]{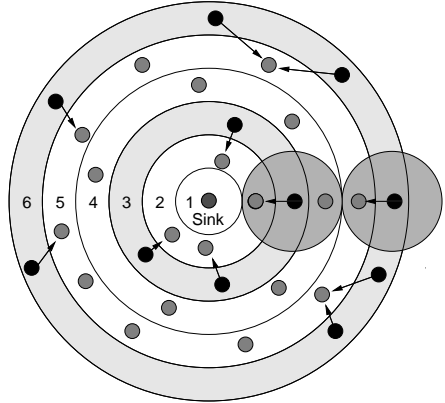}
  \caption{Network divided into different sectors, where two nodes belonging to different sectors can transmit without interfering among them.}
  \label{circle}
\end{figure}

To have the network divided into sectors allows us to coordinate the transmissions in order to reduce the hidden terminal problem explained in Section \ref{terminal}. With this information, nodes can transmit for a given period of time and do not interfere with the transmission of other sectors. In order to allow multiple sectors to transmit at the same time, these must have at least a distance of 3 between sectors \cite{gnawali2009collection}\cite{rajendran2006energy}. Therefore, if the sector $n$ is transmitting to the sectors $n\pm1$ no other sector that interferes into these sectors is allowed to transmit. However, meanwhile it would not affect that nodes in the sector $n+3$ transmit, because they will just interfere to the sectors $n+2$ and $n+4$. As an example refer to Figure \ref{circle}, where it can be observed that if nodes in sector 3 and 6 transmit at the same time no collision can happen among nodes of these sectors. Whereas, if nodes located in the sector 3 and 5 transmit simultaneously, the 
hidden terminal problem can happen in the forth sector.

Notice that HRS is not able to completely eliminate all hidden terminal problems in the network. There can be hidden terminals belonging to the same sector $n$ and therefore, they can collide when sending data to the sector $n-1$. Nevertheless, as it will be shown in Section \ref{testbed2}, dividing the network in sectors increases the end-to-end reliability as hidden terminal collisions are reduced.

\subsection{Bidirectional Link Quality Estimation (BLQE)}
\label{subsection:LQE}

For the HRS is important to determine the quality of the links, and choose the link with the greatest transmission success probability. 
In WSNs, most of the routing protocols \cite{gnawali2009collection} \cite{MultiHopLQI} \cite{tolle2005design}  prefer the LQI (Link Quality Indicator) over the RSSI (Received Signal Strength Indicator) to estimate the quality of the links because the LQI takes into account previous transmissions and computes the successful transmission probability per link. Therefore, as the LQI depends on the time and on the number of packets sent, and given that we focus on a scenario where two collections are utterly independent, there is no worth in using the LQI over the RSSI. Moreover, as it is stated in \cite{srinivasan2006rssi}, the RSSI has been under-appreciated even for values which it has shown to be very reliable. For instance, for the CC2420 chip a RSSI of $-87$ dBm assures a Packet Reception Rate (PRR) of $85\%$ \cite{srinivasan2006rssi}. The other region, from $-87$ dBm to $-94$ dBm, is called the gray area where the PRR varies radically. However, the RSSI is not a good indicator of link burstiness \cite{
srinivasan2007beta} and its degradation depending on the antenna may not follow a polynomial function of distance \cite{Rochester}, which has to be taken into account when collecting the data.

\begin{figure}[t!]
  \centering
  \epsfig{file=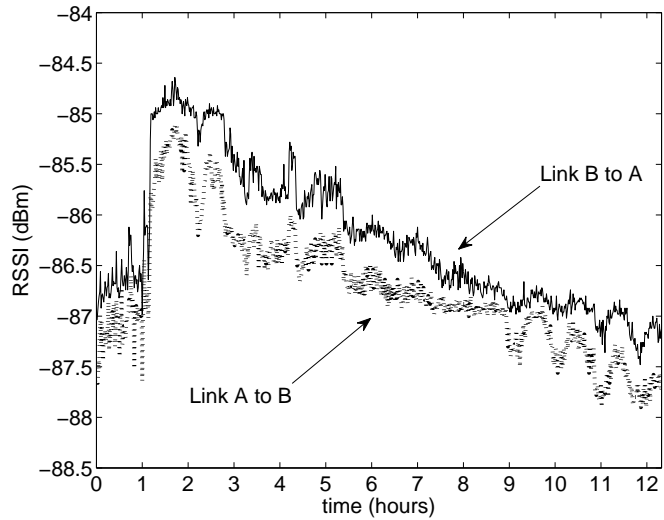,scale=0.5}
  \caption{Link symmetry evaluation between two TelosB}
  \label{symmetry}
\end{figure}

Regarding the link symmetry, and although it is well known the asymmetry of wireless links \cite{de2003performance} \cite{ganesan2002complex}, it has been shown in experimental evaluations that in some radios like the CC2420, the links are very symmetric \cite{srinivasan2006rssi}. To extend those results, we evaluated the link symmetry variation ($\gamma_{v}$) between two TelosB \cite{TelosB} (Figure \ref{symmetry}). This evaluation is done in an indoor environment for 12 hours, and shows that the link has less than $1$ dBm of difference on average between two nodes, and a total correlated fluctuation of nearly $3$ dBm.

Based on this evidence, the link quality in BLQE is calculated by evaluating if the RSSI belongs to a gray area taking into account both directions. This is done with two parameters: the gray area threshold ($\gamma_{ga}$) which defines if a link is in the gray area (useless to achieve a high packet delivery rate), and the quality threshold ($\gamma_{q}$) which assures that the link is not in the gray area for both directions. These thresholds directly depend on the transceiver used. The ($\gamma_{q}$) is calculated as follows:

\begin{center}
$\gamma_{q}$ = $\gamma_{ga}$ $- \gamma_{v}$
\end{center}

The selection of the $\gamma_{v}$ value is not straightforward, so it depends on several factors: the environmental conditions of the network, the transceiver, the transmission power and the density of nodes. Additionally, in dynamic wireless environments this parameter should be adapted depending on the varying channel conditions. As a result, the estimation of this parameter at each node would lead to an increased complexity of the protocol and energy consumption. Therefore, in this work it has been considered to fix this value for all nodes in the network to 5 dB.

In Section \ref{subsection:RHC} we have fixed a distance of 3 between sectors to avoid collisions with hidden nodes, however, this value is directly influenced by the node platform and the transmission power. As it is shown in \cite{Rochester}, where an evaluation of the RSSI behavior of the CC2420 transceiver is provided, the CC2420 is not omnidirectional, and the gray areas are affected by other factors besides of the distance. When the quality of the links is evaluated, the links in the gray areas cannot influence because they are filtered by the $\gamma_{ga}$ threshold. However, when collecting the data, if the nodes in the gray areas are not considered they can cause collisions. Hereby, the sector distance when transmitting has to be adjusted to reduce as much as possible the effect of the nodes in the gray areas.

\subsection{Randomized Transmissions}
\label{subsection:RT}

To avoid collisions when two or more nodes want to transmit at the same instant, the random MAC protocols usually wait a random backoff in order to distribute the transmissions over a certain period of time. However, using a backoff, there is a tradeoff between delay and collision probability. A larger backoff time would provide less collision probability but will compromise the transmission delay. The aim of the randomized transmissions mechanism is to spread the transmissions of the nodes that want to transmit at the same instant without modifying the MAC parameters. As an example, if there are 100 nodes that are ready to transmit, their transmissions are spread in an interval of 10 seconds, and approximately, if uniform distribution is considered, only 10 nodes will compete to access the medium during one second. Also notice that if the time to transmit data is large compared to the number of nodes, the randomized property of the random MAC protocol becomes less important. However, as the aim of the 
network is to collect the data with the minimum amount of time, the random MAC protocol is relevant. To set this transmission time, different aspects have to be taken into account: the number of nodes in the network, the MAC parameters, the transceiver characteristics and the packet lengths.

\subsection{Data Aggregation}

Most of the times the packets sent in a WSN do not achieve the maximum packet size, and usually payloads just represent a small part of the whole packet. In order to achieve better packet efficiency and consume less energy, data packets can be aggregated, which reduces transmissions and collisions. Packets have larger sizes, but overall the total amount of bytes sent is less due to the reduction of overhead.

\section{Protocol Description}
\label{description}

\subsection{Headers and addresses}

Before addressing how the protocol works, the format of the HBCP packets is introduced. HBCP has two types of packets: data and discovery. The data packets are the ones that carry the data sensed by the sensors, and the discovery packets are the ones used to notify the nodes that a collection has to be performed, create the datapaths and classify the nodes in sectors. 

\begin{figure}[ht!]
   \centering
   \subfigure[HBCP header]{\label{generalHeader}\includegraphics[width=0.5\textwidth]{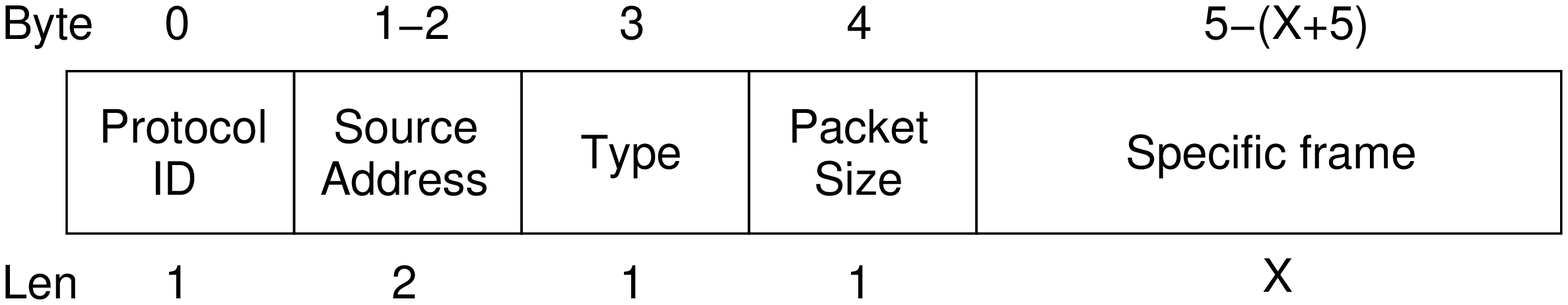}}\\
   \subfigure[Discovery frame]{\label{routingFrame} \includegraphics[width=0.5\textwidth]{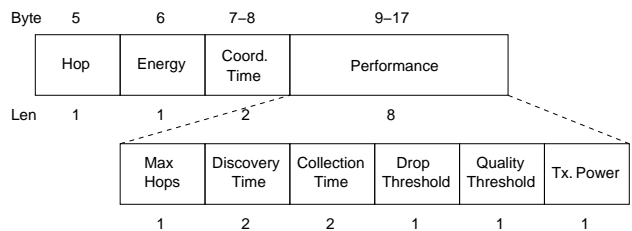}}\\
   \subfigure[Data frame]{\label{dataFrame} \includegraphics[width=0.5\textwidth]{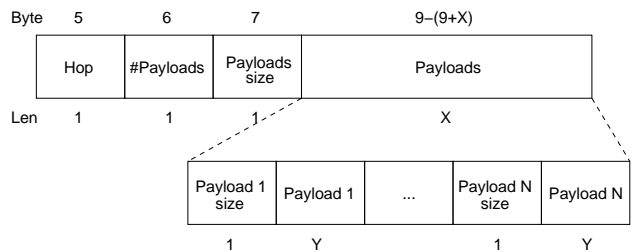}}\\
   \caption{HBCP headers. The specific frame field of the \emph{header} is where the \emph{Discovery frame} or \emph{Data frame} are placed.}
   \label{headers}
\end{figure}

Each of these two packets have a common \emph{HBCP header} (Figure \ref{generalHeader}), that is followed by the \emph{Data} or \emph{Discovery Frame} (Figures \ref{dataFrame} and \ref{routingFrame}). The HBCP header is composed by five fields: the \emph{protocol id} identifies the packet as an HBCP packet, the \emph{source address} indicates the node id, the \emph{packet size} indicates the length of the whole network packet, and the last one is the \emph{Specific Frame} which includes the \emph{Data} or the \emph{Discovery Frames}.

Regarding the \emph{Discovery Frame}, it is divided in two parts: the collection and the performance parameters. The collection parameters are three: \emph{hop}, \emph{energy level} and \emph{coordination time}, and these respectively mean the sector the node belong to, the remaining energy of a node and the coordination time that will be explained in Section \ref{subsection:TCR}. The performance parameters are \emph{max hops}, which is the maximum number of sectors that a network can have, the \emph{drop threshold} ($\gamma_{ga}$) which is the minimum quality of a discovery packet in order to be evaluated, the \emph{quality threshold} ($\gamma_{q}$) which is the indicator that bidirectional communications is assured, and the \emph{transmission power} that the devices will use. The \emph{discovery time} and the \emph{collection time} parameters will be explained in detail in Section \ref{netdisc} and \ref{collection} respectively.

The \emph{Data Frame} is simpler than the \emph{Discovery Frame}, it only contains the sector of the node that transmits the packet, the number of application data (payloads) because of data aggregation (see Section \ref{queues and forwarding}), the total size of the payload, and the sizes of each of the aggregated payloads and the total payload.  

\subsection{Network Discovery}
\label{netdisc}

This section describes how the network sectors are created and how the datapath to reach the sink is computed for each node in the network. 

\subsubsection{Transmission Range Classification}
\label{subsection:TCR}

The main goal of this function is to classify the nodes into different sectors, where each sector, as it is explained in Section \ref{subsection:RHC}, corresponds to the number of hops to reach the sink. Initially, a node does not have any kind of information about the network topology, and every time that a data collection ends, the routing information expires. The HBCP is designed for on-demand collection applications with low frequency collections, hence, as the time between two collections can be from several minutes to months, the datapaths have to be built from scratch for every collection.

The mechanism to create the routes is to coordinately send the discovery packets. The sink starts a collection sending a broadcast discovery packet that is received by the nodes that belong to sector 1. Since we have assumed that there is only one sink, just after the first transmission, every node belonging to sector 1 knows it. To discover the next level each of the nodes in sector 1 send also one discovery packet, and this process is repeated until the maximum number of hops indicated in the discovery packet is reached.

However, there are two factors that affect the network formation: 1) the collisions of the discovery packets, and 2) the capacity that a node in the sector $n$ is able to receive all the packets from the $n-1$ sector during a period of time. To alleviate the first problem nodes wait a random time, independently of their backoff, and then transmit a discovery packet (see Section \ref{subsection:RT}). This random time is between 0 and the value indicated in the discovery time field of the discovery frame. To address the second problem, nodes include in their discovery packet the random time (coordination time) they have used to transmit the discovery packet. Therefore, when a node in the sector $n$ receives a discovery packet, it knows until when it will be possible to receive more packets from the sector $n-1$. For instance, if a node receives a discovery packet at $t_0$ with a coordination time $t_c$ and a discovery time $t_d$, the node will wait until $t_0+t_d-t_c$. Regarding synchronization, it is worth to 
mention that for this type of coordination the motes clock drift is not relevant. For example, the TelosB mote at $20^\circ C$ has a large clock drift (40 ppm), what means a loss of accuracy of 2.5 ms every minute.

Regarding the random time, it is worth noticing the tradeoff between delay and collision probability. If this random time makes the network creation time to substantially increase, the environmental conditions may change, especially in volatile environments, and the path to reach the sink and sectors information may not be valid anymore. Therefore, it is key to find the right balance between collisions and reliability of paths. 

\subsubsection{Datapath Selection}
\label{datapathselection}

While the network nodes are sending the discovery packets, they use the information included in them to select which is the best next hop to reach the sink. The information used is the RSSI of the received discovery packet and the remaining energy of the parent. In the discovery packet, the criteria that the potential receivers have to follow when to select a parent (next hop) is included. The two used parameters are the $\gamma_{ga}$ and the $\gamma_{q}$. The first one is the minimum quality that a link must have, otherwise the packet will be dropped and not even considered. The second one is the threshold for a packet to be considered with good quality, hence the bidirectional communication of the link is assured.

\begin{figure}[ht!]
  \centering
    \epsfig{file=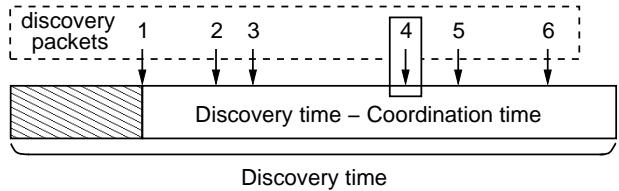,scale=0.7}
   \caption{Discovery packet reception. The node receives the first discovery packet (1), and from the information included in it knows until when it will be possible to receive discovery packets from the previous sector. In this example, the $4^{th}$ packet is the one with best quality, thus, the selected parent of the node.}
  \label{discovery}
\end{figure}

\begin{figure}[ht!]
  \centering
  \epsfig{file=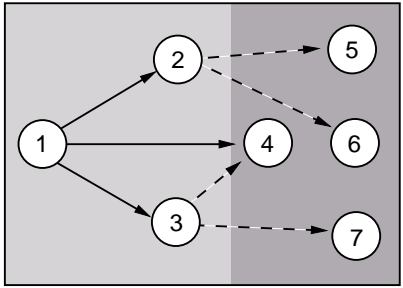,scale=0.75}
  \caption{In the first discovery period (straight line), node 4 receives a discovery packet but cannot assure bidirectional communication with node 1. Therefore, it waits for the second discovery period (dashed line) to obtain a reliable link with node 3.}
  \label{extendeddiscovery}
\end{figure}

As it is explained in Section \ref{subsection:TCR}, when a node receives a discovery packet it knows until when it would be possible to receive more discovery packets from the same sector. During this period, each node processes all the received discovery packets and picks up the one with the best quality (see Figure \ref{discovery}). However, if the quality of all the discovery packets is in between of $\gamma_{q}$ and $\gamma_{ga}$, the node will wait an additional discovery period, and try to receive a discovery packet sent from a node in the same sector it would belong to. If during this extra listening period the node receives another discovery packet with enough quality, it will take the node as a parent, otherwise, it will keep with the previous parent besides it does not have a good quality link. As an example refer to Figure \ref{extendeddiscovery}. Node 1 sends a discovery message which assures bidirectional communications between nodes 2 and 3, but not with node 4. Therefore, node 4 waits another 
discovery period to try to receive a discovery packet from nodes 2 and 3 with enough quality. To ease the understanding of this part, the parent selection process is algorithmically described in Algorithm \ref{alg}.

\begin{algorithm}[ht!]                      
\small   
\caption{Parent selection algorithm based on the received RSSI. The $current\_quality$ is the best RSSI of all the received packets, and the $backup\_parent$ the node used in case the second discovery period does not find a better parent. Packets with a quality less than $\gamma_{ga}$ are automatically discarded.}
\label{alg}
\begin{algorithmic}                   
    \State $current\_quality = -\infty ;$
    \Function{process discovery packet }{packet}
      \If{$isFirstDiscovery$}
    
	\If{$quality(packet) \ge current\_quality $}
	\State $current\_quality = quality(packet)$
    
	  \If{$quality(packet) \ge \gamma_{q}$}
	    \State $parent = packet.source\_id$
	  \EndIf
	  \State $backup\_parent = packet.source\_id$
	\EndIf
      \EndIf

    \If{$isSecondDiscovery$ }
	\If{$quality(packet) \ge \gamma_{q}$ \textbf{and} $quality(packet) \ge current\_quality $}
	  \State $parent = packet.source\_id$
	  \State $current\_quality = quality(packet)$
	\EndIf
    
    \EndIf
    
    \EndFunction
\end{algorithmic}
\end{algorithm}

The idea of taking into account the remaining energy of the nodes is to do traffic load balancing and try to avoid nodes with little remaining energy. The low energy threshold ($\gamma_{e}$) has been fixed to 15\%, but depending on the platform or the type of application this value has to be readjusted to meet the desired performance. When a node receives a discovery packet from a parent in the previous sector, it analyzes the RSSI and the battery level, and if the energy is below the fixed threshold, the link quality is readjusted to prioritize other links. Using this mechanism, HBCP intends to exploit the better datapaths until the node's energy level is critic. The link priority is done according to the quality ranges, if the RSSI is in the good quality range but with little energy, the link quality will be fixed to its minimum within the good quality range. For instance, a node has a $\gamma_{q} = K$ and receives two discovery packets with a quality of $K+5$ and $K+10$ respectively. Without taking into 
account the energy levels, the node will pick the second packet because it has better quality. However, if the battery level of the second node is critical, its quality will be downgraded to $K$. Hence, the node will pick the first packet.   

\subsection{Collection}
\label{collection}

Once a node finishes the previously explained process, it has to wait for the other nodes to start the collection. This waiting time can be calculated by computing the difference between the sector it belongs to and the total number of hops (which is included in the discovery packet), and then multiply it by the discovery time. 

\subsubsection{Data Collection Scheduling}

\begin{figure}[ht!]
  \centering
  \epsfig{file=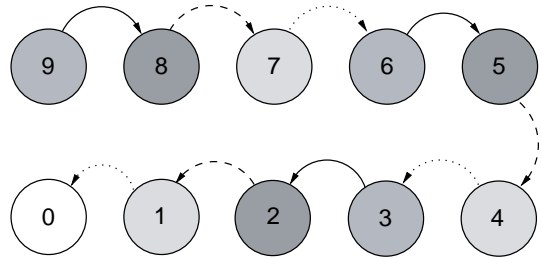,scale=0.5}
  \caption{Hop transmissions in a ten-hop network assuming a distance of 3 between sectors. In this case, in the first iteration (straight lines) sectors 3, 6 and 9 transmit at the same time to sectors 2, 5 and 8 respectively. In the second iteration (dashed lines) these sectors will transmit to the next sectors their own data and the data they received in the previous iteration. After the third iteration (dotted lines), the first iteration will start again. }
  \label{fig:hopsexample}
\end{figure}

Before to start the collection, a node knows beforehand how much time the data collection will take. It will last the total collection time, which was indicated in the discovery packet, times the number of hops. One of the first options that was considered when designing the data collection protocol was to collect the data from the outer nodes to the inner ones, but in this way the nodes would have a high probability to run out of memory: the nodes in the first sector would have to keep in memory all the network data before transmitting it to the sink. Because of this, the data collection is carried out allowing sectors that do not interfere between them to transmit at the same time. The data arrives to the sink in different waves, and the probability for the nodes to run out of memory is considerably reduced. Figure \ref{fig:hopsexample} is an example of how the protocol would behave in a network with ten hops.

\subsubsection{Data Queues and Forwarding}
\label{queues and forwarding}

As it would happen in the network discovery, if multiple nodes try to access the medium at the same time, collisions will likely occur. To reduce this problem, nodes wait a random time up to the 80\% of the data collection time. This threshold has been set to allow a node with multiple packets in the queue to have enough time to transmit them. After this initial random time a node starts the forwarding process for all the packets in the queue. 
The data aggregation is done checking how many sequential data payloads can fit into a single packet. This aggregation mechanism is not optimum to reduce the total number of transmissions, however, because its complexity is very low, it is suitable for WSNs. 

\section{Implementation in TinyOS}
\label{software}

This section presents how the HBCP architecture has been implemented in TinyOS 2.1\footnote{Code available at: http://code.google.com/p/hbcp/}. 

\begin{figure}[ht]
  \centering
  \small
  \epsfig{file=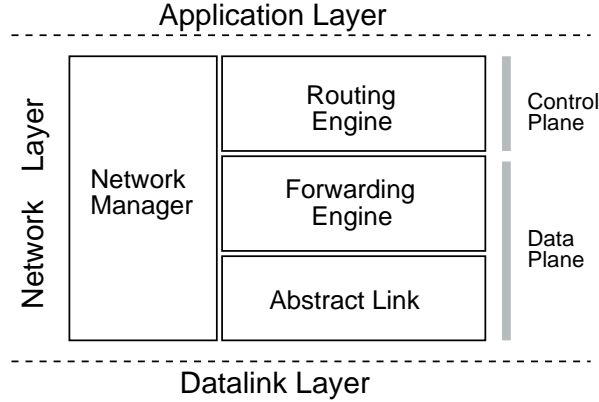,scale=0.7}
  \caption{HBCP TinyOS components}
  \label{layers}
\end{figure}

Figure \ref{layers} shows the different HBCP components implemented in TinyOS. The Routing Engine is the component responsible for analyzing the received discovery packets, choose the best node as a parent and store the datapath information. In the Forwarding Engine there are the queues that control the incoming and the outgoing packets, and it is in charge of dispatching the packets to the different components. Finally, the Network Manager is the component in charge of controlling the behavior of the other components depending on the node state. The role of the Network Manager is crucial due to the number of concurrent timers in the code and because each component can behave in a different way depending on the node state. For example, the Routing Engine will not be allowed to analyze new received discovery packets during the data collection. 

Additionally, the Network Manager is the interface between the application and the Network Layer. Therefore, it is in charge of notifying the Application Layer that a collection is about to start, so it can obtain the value from its sensor and forward it to the Network Layer.


\section{Performance Evaluation}
\label{evaluation}

This section evaluates the HBCP performance, with special focus on the behavior of the new proposed mechanisms in a real environment.
The implementation has been evaluated using the TelosB platform, with the CSMA link layer without Low Power Listening (LPL) \cite{moss2007low} active. LPL has not been used because it is not suitable for networks that demand a great number of transmissions during a short period of time, i.e, nodes are intended to work intensively during few seconds, and afterward turn into sleep mode until the next collection. 

\begin{table}[ht!]
  \begin{center}
    \begin{tabular}{|l|c|}
      \hline
      Parameter & Value\\
      \hline
      Max Hops & 10 \\
      Discovery time & $1000$ ms\\
      Collection time & $2000$ ms \\
      $\gamma_{ga}$ & $-87$ dBm \\
      $\gamma_{q}$ & $-82$ dBm\\
      Transmission power & $0$ dBm \\
      \hline

    \end{tabular}
  \end{center}
  \caption{Parameters considered in the Testbed}
  \label{table:testbedparameters}
\end{table}

To evaluate the two mechanisms, two testbeds are deployed. One for the evaluation of the datapath creation and data collection, which consisted of placing 30 nodes randomly throughout a three storey building, and checking during 24 hours how the PRR behaves for the next hop and end-to-end. On the second testbed, the performance of the data collection with different collection times is evaluated, as well as how the protocol would behave if HRS is not applied. One hundred collections are performed with and without HRS, with the collection slot times set to 250, 500, 750, 1000, 1500 and 2000 ms.

Taking as a reference the previous work on the CC2420 chip presented in \cite{srinivasan2006rssi}, and assuming a $\gamma_{v}$ of $5$ dB, the $\gamma_{q}$ and the $\gamma_{ga}$ were fixed to $-82$ and $-87$ dBm respectively. For the two testbeds  the transmission power is fixed to $0$ dBm.  Obviously, those values can vary depending on the transceiver and the transmission power of each node in the network.  

The discovery time parameter is fixed to 1000 ms for the two testbeds, as this value grants enough time for the nodes to receive at least one discovery packet. The largest collection time in the first testbed (2000 ms) is established by checking that the sink can receive all the packets in one collection. Moreover, in the second testbed, where the collection time parameter is evaluated, it is shown that 2000 ms is enough time to avoid collision between nodes of the same sector.  

Regarding the data link layer, the number of retransmissions is fixed taking as reference the policy used in CTP \cite{gnawali2009collection}, which is up to 32 retransmissions per packet. 

\subsection{Testbed 1: Network formation}

\begin{figure}[t!]
  \centering
  \epsfig{file=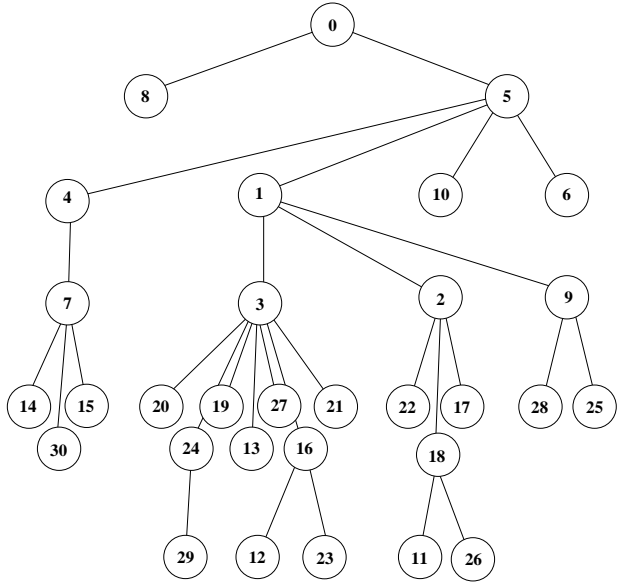,scale=0.4}
  \caption{Most frequent tree created during the testbed.}
  \label{sc}
\end{figure}

\begin{figure}[ht!]
    \centering
    \subfigure[Average number of hops per node. On average, in the 99.98\% of the collections, a node belonged to the same sector.]		{\label{ea}\epsfig{file=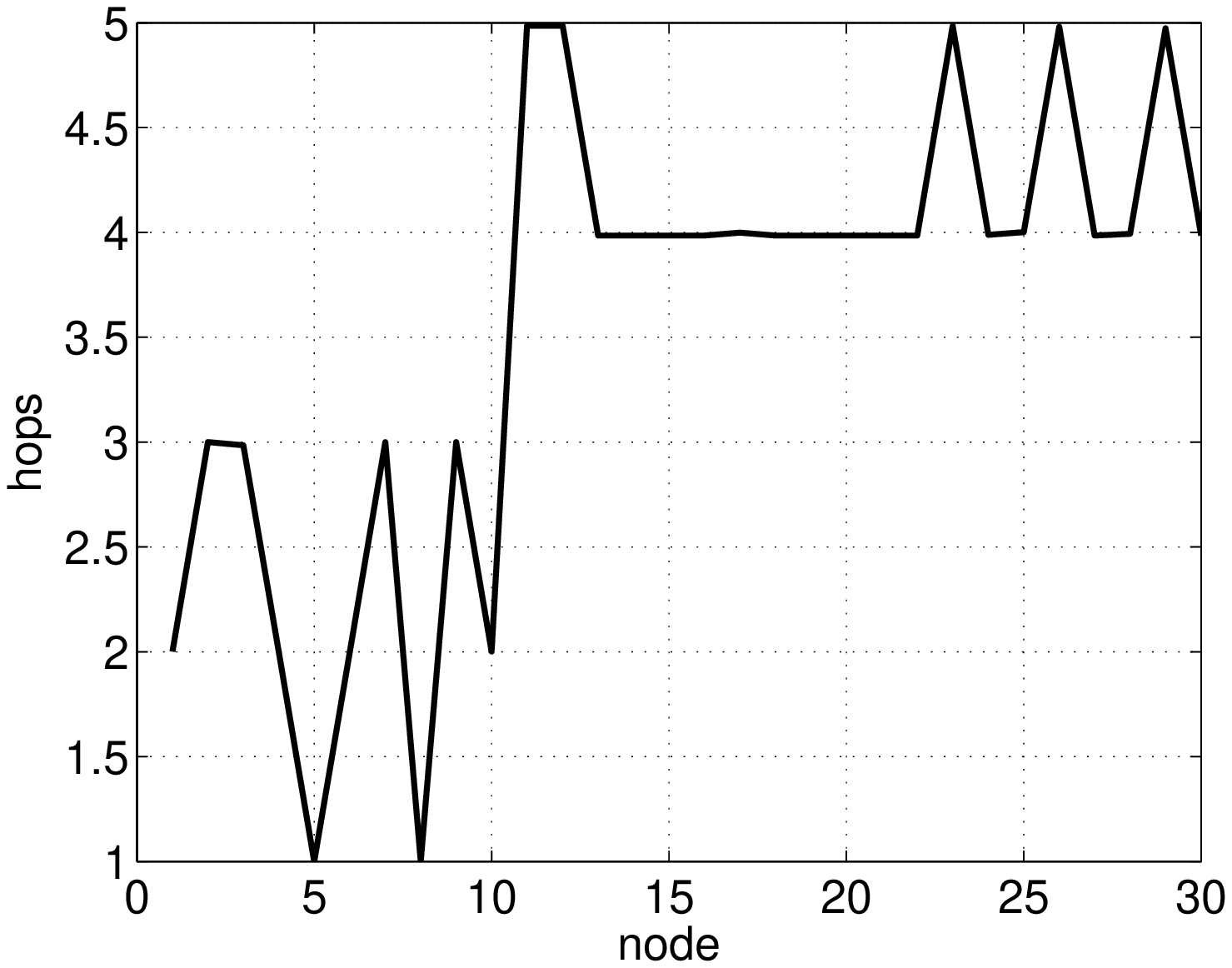,scale=0.45}}
    \subfigure[Percentage of times the most used parent was elected for each node]{\label{ec}\epsfig{file=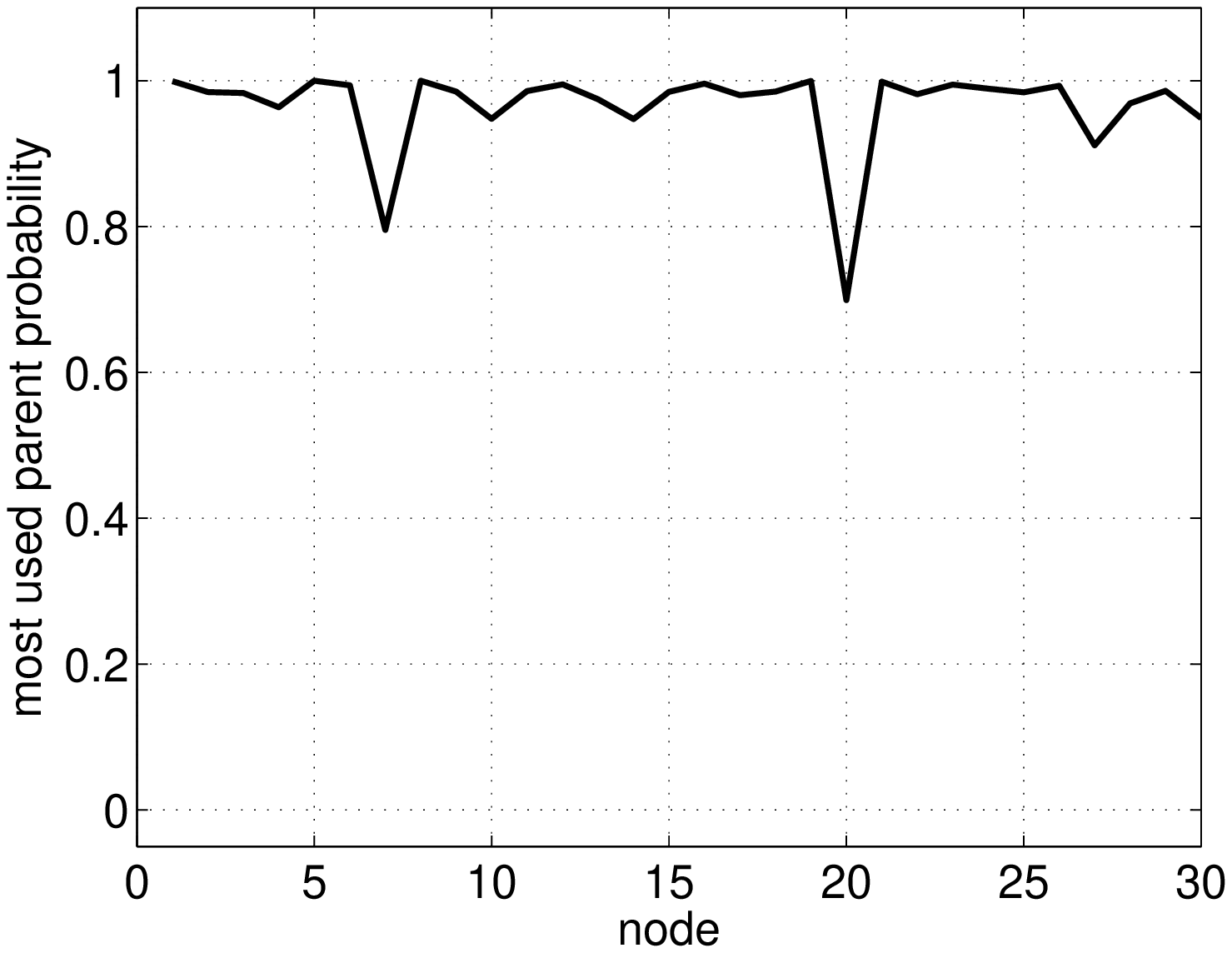,scale=0.45}}
    \subfigure[Average RSSI of the received discovery packet. The links are shown in Figure \ref{sc}.]{\label{eb}\epsfig{file=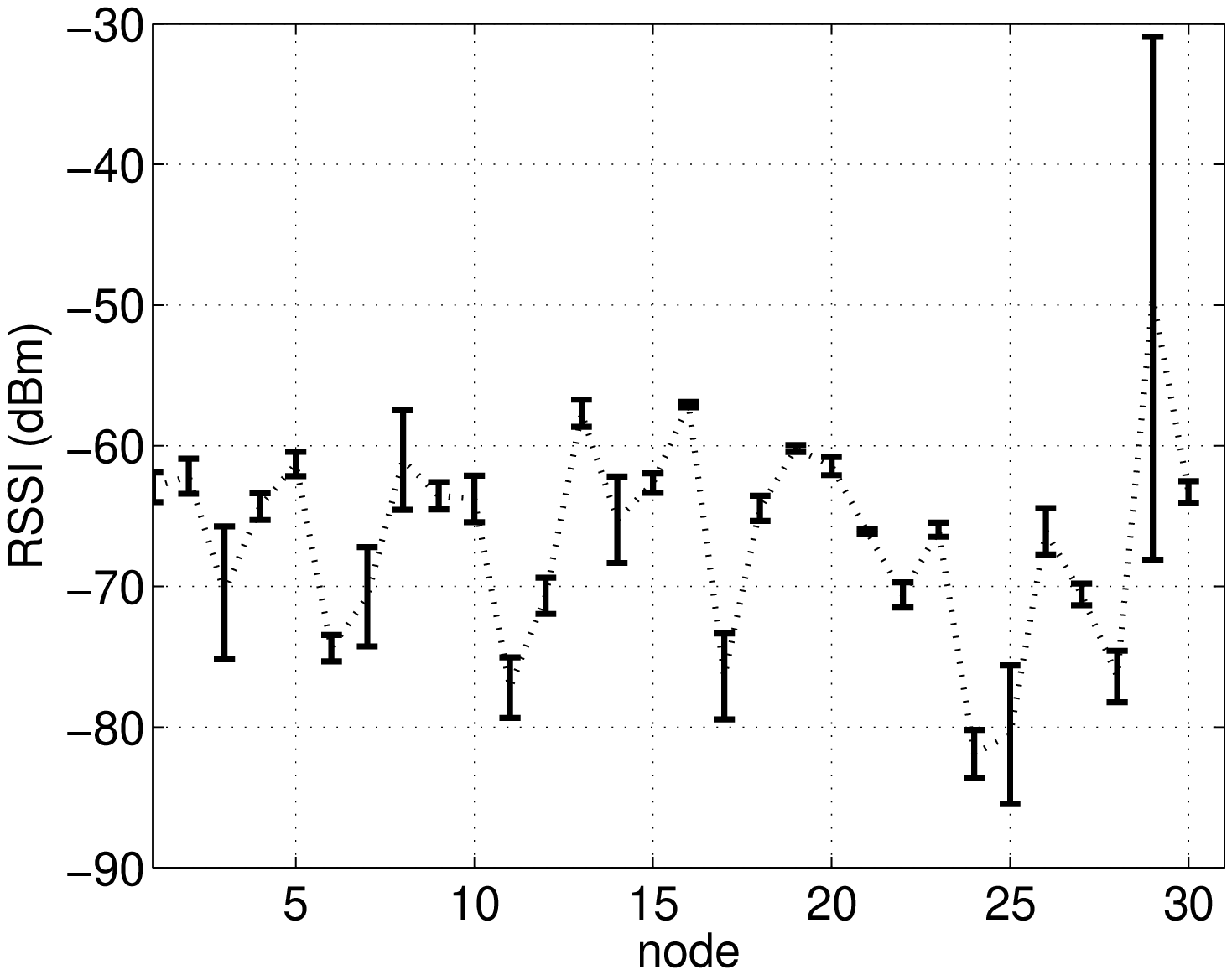,scale=0.45}}\\
    \caption{Datapath metrics of the first testbed.}
    \label{datapathfig}
\end{figure}

\begin{figure*}[ht!]
    \centering
    \subfigure[End-to-end reliability per node. It only includes the application packets generated by the node. (the dotted line is the average)]{\label{ed}\epsfig{file=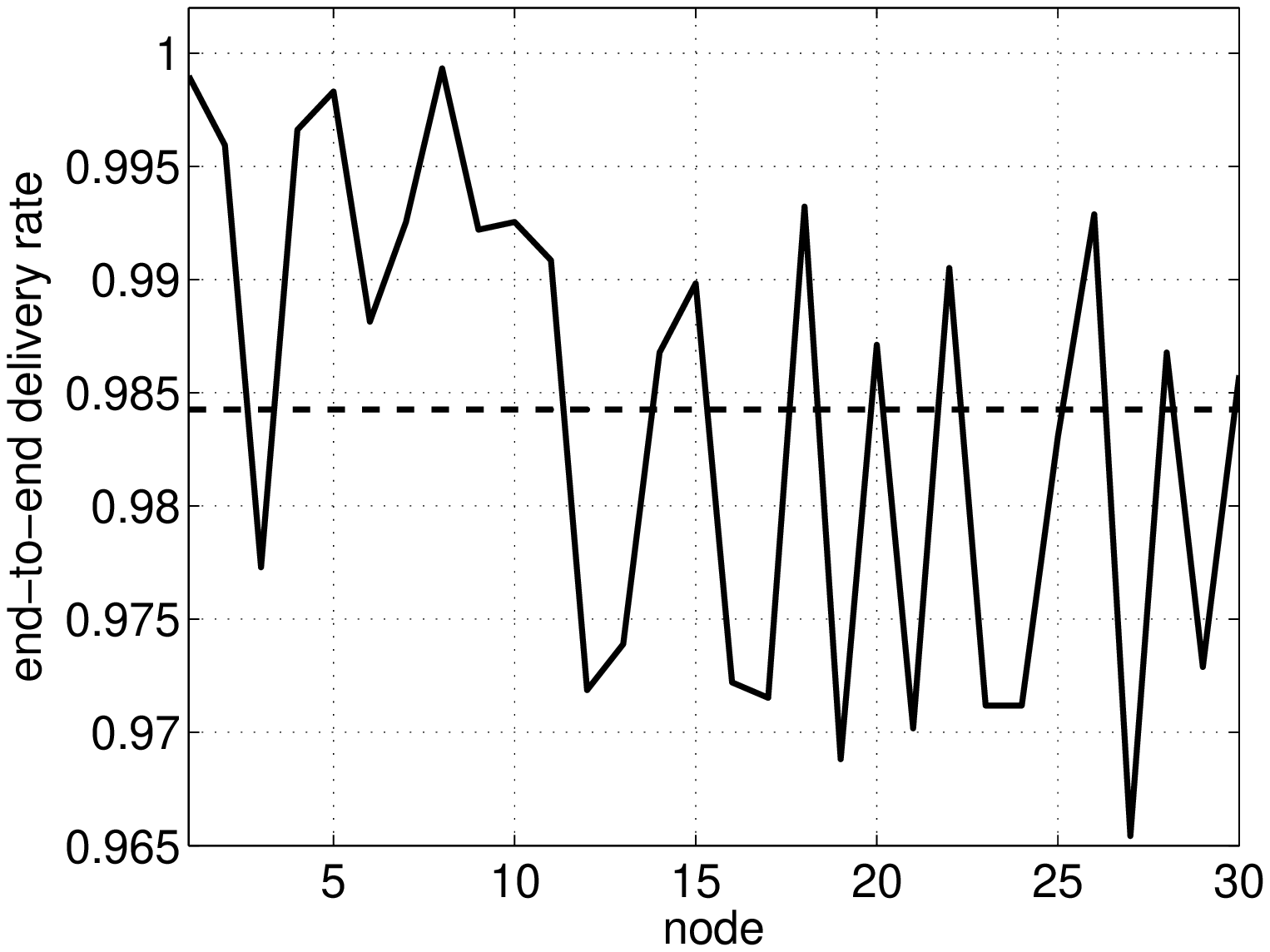,scale=0.45}}  \footnotesize
    \subfigure[Number of packets acknowledged per node. It includes the applications packets generated by the node and the packet forwarded from other nodes (the dotted line is the average).]{\label{ee}\epsfig{file=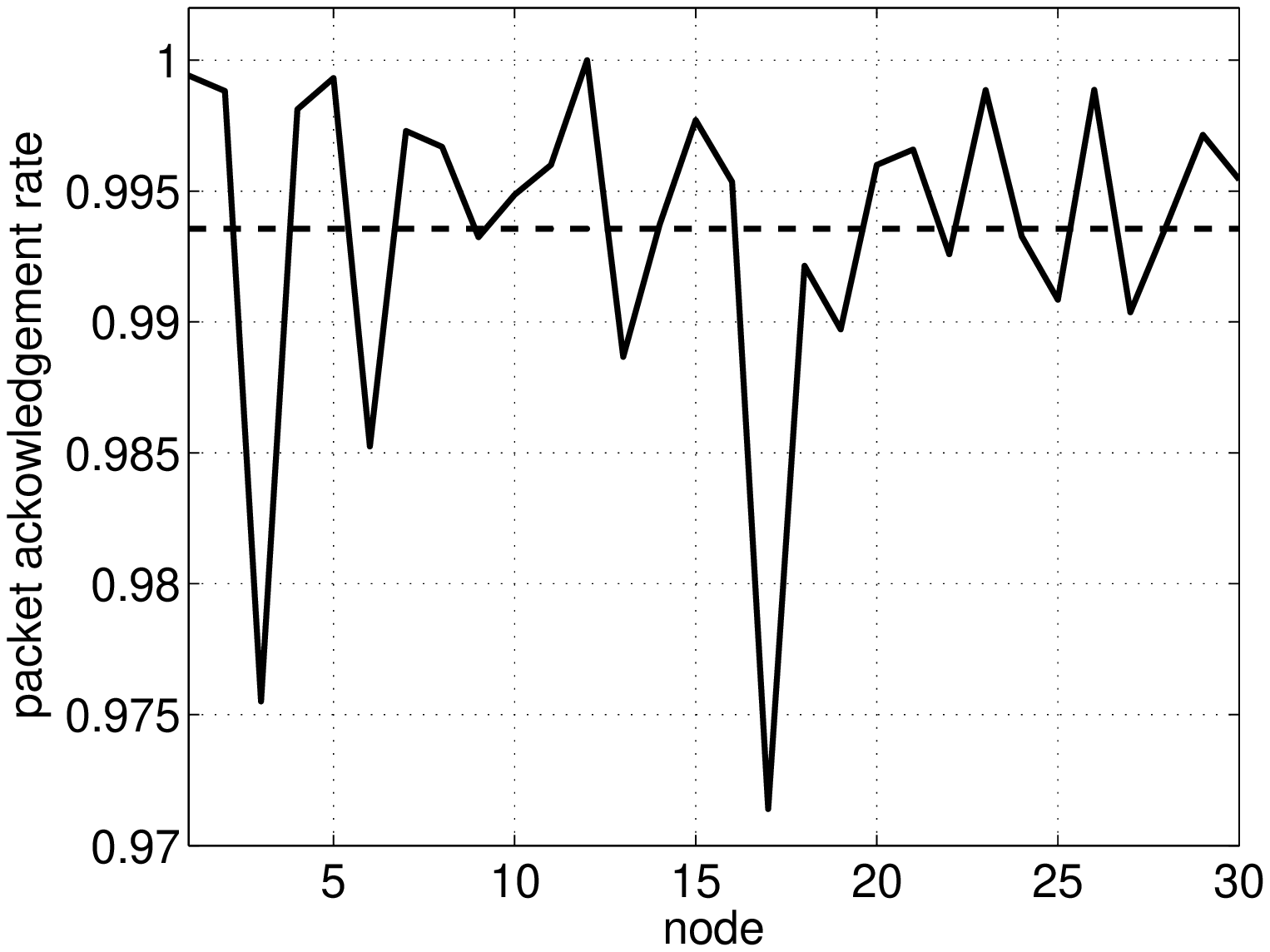,scale=0.45}}\\
    \subfigure[Number of packets transmitted per node per collection]{\label{ef}\epsfig{file=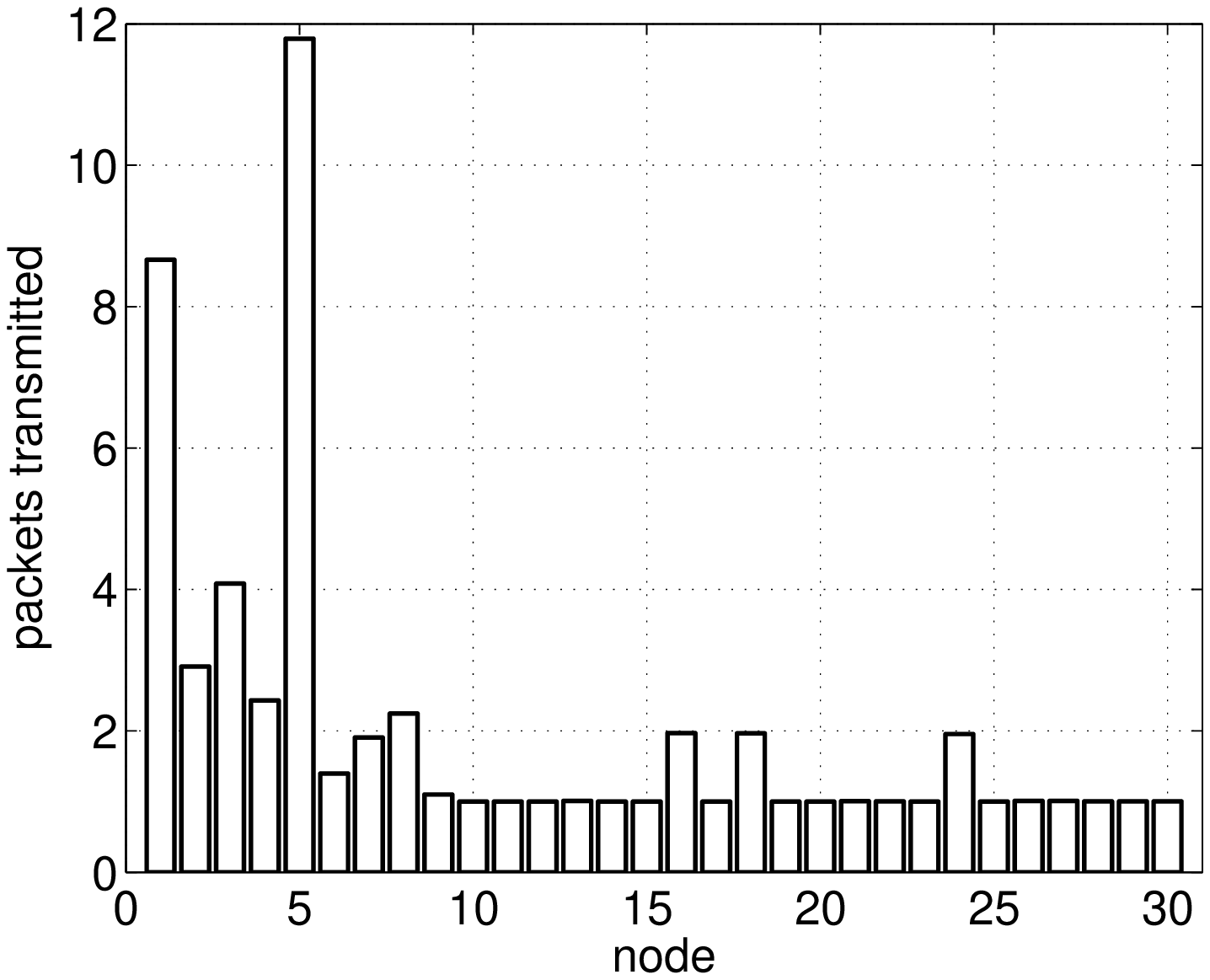,scale=0.45}}
    \caption{Transmission metrics of the first testbed.}
    \label{collectionfig}
\end{figure*}

In all the collections, the tree that has been formed most of the times is the one shown in Figure \ref{sc}. In Figure \ref{ea} is depicted the average number of hops per node for all the collections,  where despite some changes, the nodes nearly always (99.98\%) tend to  belong to the same sector. However, it does not happen the same with the selected parent in each collection, as it is shown in Figure \ref{ec}. For the majority of the collections, most of the nodes tend to chose the same parent, but it is not as steady as the number of hops shown in Figure \ref{ea}. There is a specific case that is interesting to highlight, and it is the case of nodes 7 and 20, which drastically differentiate from the others because of having a most used parent rate of 80\% and 70\% respectively. Although this, the change of parent variation does not affect the node performance in terms of end-to-end delivery (Figure \ref{ed}) and next-hop acknowledgment rate (Figure \ref{ee}). The case of node number 20 is quite 
particular, because as it shown in Figure \ref{eb}, where the average RSSI per node is depicted, the variance of this node is slightly over 1 dB and suggests that the nodes that try to be its parent have somehow similar characteristics. In the case of node 7, the cause of this variation is not straightforward and might be due to environmental changes during the testbed, like doors openings or people moving.

Regarding the collection reliability, it is interesting to check the end-to-end delivery and next-hop delivery rates, and how those vary depending on the sector the nodes belong to. In the next hop delivery rate, the number of packets acknowledged per node is checked, and as it is expected the next hop link quality is independent of the sector the node belongs to (Figure \ref{ee}), with an average over 99\%. However, the number of hops matter in the end-to-end delivery rate. If  Figures \ref{ea} and \ref{ed} are compared, it can be observed how the nodes that belong to further sectors have lower end-to-end delivery rates. Nonetheless, the overall end-to-end delivery rate is considerably high, with an average of 98.4\%.

The last thing to check in this testbed is to see how data aggregation benefits in terms of reducing the number of transmissions per node. In Figure \ref{ef} the average transmitted packets per collection and node are shown, and as it can be presumably deduced from Figure \ref{sc}, nodes 1 and 5 are the ones that transmit a greater number of packets per collection, with an average of 8.66 and 11.79 respectively. In this scenario, due to the payload size, the maximum packet aggregation was set to 3. The total average number of packets transmitted per collection was 60.43, and compared to the 107 transmitted packets that would have been expected without HRS, it is a reduction of over 40\%. However, this results are only valid for this case, hence the transmitted packet reduction highly depends on the number of hops and the location of the nodes within the network. It must also be taken into account that if smaller application packets were used, the number of transmissions could have been reduced even more. 

From Table \ref{tableresults} similar conclusions can be obtained, but instead of analyzing the results for every node, the performance of each sector is analyzed. As it was previously observed, the end-to-end delivery rate decreases with the number of hops. But there is an exception, and it is in $5^{th}$ sector, where its delivery rate is better than in the $4^{th}$ sector. Nonetheless, the variation is quite small and it is due to the fact that the nodes in the $5^{th}$ sector are not uniformly distributed along the nodes in the $4^{th}$ sector. Regarding the acknowledgment rate, in all sectors it is over 99\% and has some sort of correlation with the density of nodes in each sector. 

\begin{table}[ht!]
  \centering
  \small
  \begin{tabular}{|l |c c c c c|}
    \hline
    & \multicolumn{5}{c|}{Sectors}\\
    \hline
    Parameters & 1 & 2 & 3 & 4 & 5\\
    \hline
    Number of nodes &  2 & 4 & 4 & 15 & 5\\
    \hline
    Packets sent per coll. & 14.03 & 13.48 & 9.99 & 17.91 & 5\\
    \hline
    Exp. pkt. sent without HRS & 30 & 28 & 24 & 20 & 5\\
    \hline
    End-to-end delivery rate & 99.88 & 99.41 & 98.95 & 97.53 & 97.99\\
    \hline 
    Acknowledgment rate & 99.8 & 99.44 & 99.12 & 99.18 & 99.82\\
    \hline
 \end{tabular}
 \caption{Sectors performance of the randomly deployed testbed}
 \label{tableresults}
\end{table}

\subsection{Testbed 2: Collection times}
\label{testbed2}

In the second testbed, the effect of different collection times is evaluated. The aim is to check how the HBCP responds to this parameter, and to check whether applying HRS adds any benefit to the end-to-end delivery rate. Somehow, through varying the collection time per sector it is possible to achieve an approximation of how the network would behave with a higher density of nodes. As it was explained in Section \ref{subsection:TCR}, the collection slot time has to be large enough to allow all the nodes in a sector to transmit, but with the drawback that a large value may make the routes expire, especially in volatile environments. In addition, the protocol must perform the collection as fast as possible so the nodes can go back to sleep and save energy. 

Two different topologies are deployed, the former with a bottle-neck: one node in the first sector, another in the second sector, and twenty-three in the third sector (1-1-23). In the latter nodes are uniformly distributed in each sector (10-10-10). In the collection mechanism without HRS a node can transmit at any instant within the whole collection time of all sectors. 

\begin{figure}
  \centering
    \epsfig{file=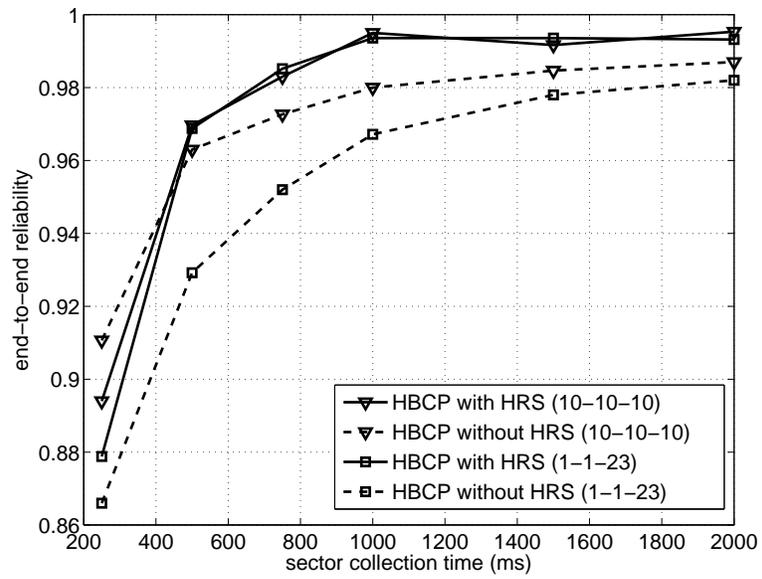,scale=0.5}
  \caption{HBCP end-to-end delivery rate with and without HRS for different collection times.}
  \label{stress}
\end{figure}

The obtained results are plotted in Figure \ref{stress}, and clearly show that nodes that implement the HRS tend to converge more rapidly to its maximum end-to-end delivery rate. Overall, using HRS is always better except in the case where the nodes are uniformly distributed and the collection time is 250 ms. In this case, with HRS the number of collisions between nodes in the same sector is greater than the collisions without HRS. However, in the bottle-neck topology it happens exactly the opposite, especially when the collection time ranges from 500 to 1500 ms. During this 1000 ms span, as nodes in each sector have enough time to transmit, the collisions between nodes of a same sector are less than the ones of nodes from different sectors.

It is worth to notice that HRS reduces the number of collisions between nodes in different sectors, but it increases the collisions of nodes in the same sector. As the nodes in the same sector have to transmit during a limited period of time, it is more likely that a node receives two packets from two different nodes that cannot overhear each other. In contrast, this is less likely without HRS, but there is no kind of control of the transmissions of nodes from different sectors.

\clearpage


\section{Related Work} 
\label{Sec:RelWork}

Currently, there are several developed and implemented network protocols in WSNs, however, most of which have been designed to send event-based data rather than answering to explicit queries from the sink. In this section we present three protocols: LEACH \cite{heinzelman2000energy}, CTP \cite{gnawali2009collection} and Direct Diffusion \cite{intanagonwiwat2003directed}, which have some similarities with the HBCP. Furthermore, it has been included some mechanisms that, although very different in nature, aim to achieve the same goal as the HRS and BLQE.

LEACH is a low-energy, self-organizing, adaptive clustering routing protocol that uses randomization to distribute the energy consumption evenly among the sensors in the network. It belongs to the hierarchical family of network protocols, where the different sensors in the network are divided in clusters, and its main purpose is to increase the whole sensor network life-time through network traffic balancing. Sensors elect themselves as cluster-heads at a given time with a certain probability, and the other nodes select from the existing clusters which they want to belong to by choosing the cluster-heads that has the minimum delivery cost (based on the RSSI). Once the network is configured, a cluster-head creates a schedule for each node in the cluster, similarly to TDMA, so the non-cluster-heads can turn off the radio except when they have to transmit, which minimizes the energy dissipated per sensor. LEACH does not define how data has to be collected, however, in contrast to HBCP, nodes always have a route 
to reach the sink. In addition, one of the drawbacks of LEACH is that the network cannot have more than two hops, therefore, it is not suitable for large multihop networks. 

The Collection Tree Protocol (CTP) is a collection protocol based on events designed to improve the data collection reliability while sending a reduced number of control packets (beacons). It has been implemented to be hardware and MAC protocol independent, and the protocol evaluation in different testbeds and platforms has shown a reliability results between 90-99.9\%, while sending up to 73\% fewer control packets than existing approaches \cite{MultiHopLQI}. CTP is basically characterized by the way routes are computed, and how beaconing is adapted to do not overload the network. Routes are address-free and the nodes just know its next hop and the cost to reach the sink. The parent selection mechanism is done using the ETX (Expected Transmission) \cite{couto2005high} value which is computed according to the route cost and the parent reliability. Firstly, the sink starts sending a beacon with an ETX equal to zero, the nodes that receive that beacon announce their cost to reach the sink sending another 
beacon, and that is 
repeated for the following levels on the network. In case that a node receives more than one packet it will always choose the one with the lowest ETX as it would be the best route to reach the sink. With this routing algorithm a node knows if there is a loop in the network just comparing the ETX values, because the parent ETX should always be lower than the one of its son. The routes are periodically transmitted, however depending on the network conditions, the routes maintenance frequency decreases, and only increases if it is needed (link failure). Differently from the HBCP, CTP is thought to collect data with high frequency. Moreover, as the data transmissions are based on events, there is no kind of coordination, so nodes should always be awake to be able to receive data to later forward it. Because of this, CTP is not energy-efficient for low periodicity data collections. 

Direct Diffusion is a data-centric routing protocol for on-demand collections, where the sink broadcast a query to the whole network which contains parameters related to the type of information it is looking for, such as name of objects, interval, geographical area, interval, etc. When nodes receive a query, they analyze it, and if their information matches with the sink interests they start sending data to the sink for a given period of time, which was fixed in the query. Despite the fact that HBCP and Direct Diffusion are on-demand protocols, their purposes are very different. Direct Diffusion is designed for event-based applications that require to receive data from a portion or the entire network for a given interval of time, while HBCP is designed to collect one reading from each sensor in the network. In addition, Direct Diffusion does not implement any mechanism to reduce the collisions due to hidden nodes, what makes it not suitable for collecting a large amount of data during a short period of time. 

There are some TDMA-like protocols (as the LMAC defined by Van Hoesel et. al) \cite{van2004lightweight} in which each sensor node picks a random slot from the free slots (slots not used in a 2-hop distance) in order to avoid hidden terminal interference. In LMAC sensor nodes exchange information about the slots they see as free. By doing this, a sensor node can select a slot from the 2-hops away unused slots, i.e., a slot considered free by all of its neighbors. The slot selection starts by a message sent by the sink. After receiving it, sensor nodes select the slot to use for transmitting data. Data transmissions are preceded by control messages that serve to synchronize sensor nodes and to inform about the intended destination of the data packet, then non interested nodes can go to sleep. The LMAC concept of reusing slots is very close the HRS concept of allowing nodes in different sectors to transmit during a given period of time. Moreover, the reuse distance (slot/sector) used in the LMAC and the HRS is 
the same. However, given that LMAC needs to synchronize every node in the network, the configuration process is more complex and requires more transmissions of control messages, and as a result would entail higher energy consumption.

The link quality estimation has gained attention in the last years because of the impact that links have in the delivery rate and energy consumption metrics. The basic metrics used for estimating the quality of a link are three: RSSI, LQI and PRR. From these, the two most used are the RSSI and LQI because they have shown that for certain thresholds they are capable of providing 95\% reliability \cite{srinivasan2006rssi}. However, differently from the RSSI, the LQI quality estimation is more accurate but needs at least 120 packets for a good estimation. Overall the different Link Quality Estimators (LQEs), the Fuzzy Link Quality Estimator (F-LQE)\cite{baccour2010testbed}, by Baccour et al., has shown to outperform most of the LQE -based estimators \cite{couto2005high}\cite{woo2003evaluation}\cite{fonseca2007four}\cite{cerpa2005temporal}. F-LQE does not only take into account one link metric but four: the smoothed packet reception rate, the asymmetry level, the average signal to noise ratio and the link 
stability. However, the penalty of obtaining such an accurate metric is time, which makes the F-LQE not suitable for volatile environments where a fast link estimation is required. In order to reduce the time required to estimate a link quality, Boano et. al. presented the Triangle metric \cite{boano2010triangle}. This LQE metric exploits the correlation between the LQI, PRR and SNR metrics to obtain a link estimation in a fast way. Results show that a good link estimation can be achieved with just 10 packets, and that the accuracy of the link estimation increases with the number of packets sent. Nevertheless, besides reducing the number of packets to estimate the quality of a link, the total number of control packets grows exponentially with the number of nodes. With BLQE, as we have shown in Section \ref{evaluation}, it is possible to obtain a good estimation of the quality of a link using only a reduced number of control packets (linearly dependent on the number of nodes). An approach that only uses the 
RSSI metric is the Kalman filter based link quality estimator (KLE) \cite{senel2007kalman}, which estimates the quality of a link with a single packet. With the RSSI of the received packet and the ground noise, the KLE does an approximation of the SNR and the PRR of the link. The main drawback of the KLE is that the function that maps the Received Signal Strength (RSS) with the SNR does not adapt to dynamic environments. BLQE also suffers in dynamic environments, however, the $\gamma_v$ parameter is easier to compute and it obtains the link quality in both directions.
   

 
\section {Conclusions}
\label{conclusions}

In this paper we have presented the Hierarchical-Based Collection Protocol for on-demand data collections in WSNs, which main features are the HRS and BLQE mechanisms. The first one estimates the link quality between two nodes based on the RSSI of the received discovery packets, and the second one organizes the nodes in different sectors, in a way that when performing the data collection the hidden terminal problem is reduced. Both mechanisms, along with the other mechanisms included in the HBCP, have been implemented in TinyOS 2.1, and have been experimentally evaluated using two testbeds. From the results obtained from the testbeds, we have shown that the BLQE mechanism is able to provide good quality links achieving a reliability over $99$\%. Regarding the HRS, its benefits are especially appreciated in networks with a high number of hops and bottle-neck topologies. 
In addition to the main two mechanisms, we have also observed that data aggregation can substantially reduce the number of packets transmitted and therefore, reduce the number of collisions. However, its benefits highly depend on the type of topology and the maximum aggregation per packet which directly depends on the size of the payloads. 

One of the drawbacks of just using one packet per node to discover the network is that a node may not receive any discovery packet due to collisions or channel outages. In this case, the node will not send data to the sink because it will not be aware that a data collection has been started. How the discovery time affects the number of collisions, and which is the probability that a node does not receive a discovery packet is an interesting point to be addressed in future work. The discovery time is crucial to ensure that all nodes participate in the collection and to reduce the amount of time the nodes are awake. In addition, it would be worthwhile to reduce the impact of a link break-down, especially in nodes in sectors near to the sink, which have to forward multiple data packets. A node could store multiple parents addresses or send broadcast data packets to nodes in the next sector.

Finally, if very large networks are considered, it would be worth to use multiple sinks within the same network. This will decrease the number of hops and also reduce the total collection time. The protocol presented in this work can be easily adapted to work with multiple sinks if the queries sent are synchronized. Therefore, and according to how datapaths are created in the HBCP, only the nodes at the same distance (in hops) from two or more sinks will receive queries from different sinks. The other nodes will only receive queries from one sink, and will not even notice of the existence of other sinks in the network. Furthermore, independently of the number of queries a node receives, data will only be forwarded to one sink. One of the problems of using multiple sinks is that the nodes at the same distance from different sinks will receive a higher number of queries, which makes collisions more likely to happen if the discovery time is static. Because of this, in future work we would like to investigate 
how the discovery time can be dynamically adapted depending on the network conditions.



\bibliographystyle{unsrt}
\bibliography{hbcp}

\end{document}